\documentclass[10pt,aps,prd,superscriptaddress,floatfix,nofootinbib]{revtex4}

\usepackage[dvips]{graphicx}
\usepackage{xspace}
\usepackage{latexsym}
\usepackage{amssymb}
\usepackage{times}
\usepackage{mathptm}

\usepackage{nopageno}

\begin{document}
\title{The Hyper-Kamiokande Experiment - Snowmass LOI}

\newcommand{\UCI}{\affiliation{University of California, Irvine, Department of Physics and Astronomy, Irvine, California, USA}}
\newcommand{\SACLAY}{\affiliation{IRFU, CEA, Universit\`e Paris-Saclay, Gif-sur-Yvette, France}}
\newcommand{\CHONNAM}{\affiliation{Chonnam National University, Department of Physics, Gwangju, Korea}}
\newcommand{\DONGSHIN}{\affiliation{Dongshin University, Laboratory for High Energy Physics, Naju, Korea}}
\newcommand{\LLR}{\affiliation{Ecole Polytechnique, IN2P3-CNRS, Laboratoire Leprince-Ringuet, Palaiseau, France}}
\newcommand{\LPNHE}{\affiliation{Laboratoire de Physique Nucleaire et de Hautes Energie, IN2P3/CNRS, Sorbonne Universit\`e, Paris, France}}
\newcommand{\EDINBURGH}{\affiliation{University of Edinburgh, School of Physics and Astronomy, Edinburgh, United Kingdom}}
\newcommand{\GENEVA}{\affiliation{University of Geneva, Section de Physique, DPNC, Geneva, Switzerland}}
\newcommand{\GIST}{\affiliation{GIST College, Gwangju Institute of Science and Technology, Gwangju, Korea}}
\newcommand{\IMPERIAL}{\affiliation{Imperial College London, Department of Physics, London, United Kingdom}}
\newcommand{\BARI}{\affiliation{INFN Sezione di Bari and Universit\`a e Politecnico di Bari,Bari Italy}}
\newcommand{\NAPOLI}{\affiliation{INFN Sezione di Napoli and Universit\`a Federico II di Napoli, Dipartimento di Fisica, Napoli, Italy}}
\newcommand{\PADOVA}{\affiliation{INFN Sezione di Padova and Universit\`a di Padova, Dipartimento di Fisica, Padova, Italy}}
\newcommand{\ROME}{\affiliation{INFN Sezione di Roma, Universit\`a Sapienza, Dipartimento di Fisica, Roma, Italy}}
\newcommand{\INR}{\affiliation{Institute for Nuclear Research of the Russian Academy of Sciences, Moscow, Russia}}
\newcommand{\KEK}{\affiliation{High Energy Accelerator Research Organization (KEK), Tsukuba, Japan}}
\newcommand{\JPARC}{\affiliation{J-PARC Center, Tokai, Japan}}
\newcommand{\SOKENDAI}{\affiliation{SOKENDAI (The Graduate University for Advanced Studies), Tokai, Japan}}
\newcommand{\KOBE}{\affiliation{Kobe University, Department of Physics, Kobe, Japan}}
\newcommand{\KYOTO}{\affiliation{Kyoto University, Department of Physics, Kyoto, Japan}}
\newcommand{\LNF}{\affiliation{Laboratori Nazionali di Frascati, Frascati, Italy}}
\newcommand{\LANCASTER}{\affiliation{Lancaster University, Physics Department, Lancaster, United Kingdom}}
\newcommand{\LIVERPOOL}{\affiliation{University of Liverpool, Department of Physics, Liverpool, United Kingdom}}
\newcommand{\LSU}{\affiliation{Louisiana State University, Department of Physics and Astronomy, Baton Rouge, Louisiana, USA }}
\newcommand{\MADRID}{\affiliation{University Autonoma Madrid, Department of Theoretical Physics, Madrid, Spain}}
\newcommand{\MIYAGI}{\affiliation{Miyagi University of Education, Department of Physics, Sendai, Japan}}
\newcommand{\NAGOYA}{\affiliation{Nagoya University, Graduate School of Science, Nagoya, Japan}}
\newcommand{\KMI}{\affiliation{Nagoya University, Kobayashi-Maskawa Institute for the Origin of Particles and the Universe, Nagoya, Japan}}
\newcommand{\STELAB}{\affiliation{Nagoya University, Institute for Space-Earth Environmental Research, Nagoya, Japan}}
\newcommand{\NCBJ}{\affiliation{National Centre for Nuclear Research, Warsaw, Poland}}
\newcommand{\OKAYAMA}{\affiliation{Okayama University, Department of Physics, Okayama, Japan}}
\newcommand{\OCU}{\affiliation{Osaka City University, Department of Physics, Osaka, Japan}}
\newcommand{\OXFORD}{\affiliation{Oxford University, Department of Physics, Oxford, United Kingdom}}
\newcommand{\RAL}{\affiliation{STFC, Rutherford Appleton Laboratory, Harwell Oxford, and Daresbury Laboratory, Warrington, United Kingdom}}
\newcommand{\REGINA}{\affiliation{University of Regina, Department of Physics, Regina, Saskatchewan, Canada}}
\newcommand{\RIO}{\affiliation{Pontif{\'\i}cia Universidade Cat{\'o}lica do Rio de Janeiro, Departamento de F\'{\i}sica, Rio de Janeiro, Brazil}}
\newcommand{\RWTH}{\affiliation{RWTH Aachen University, III. Physikalisches Institut, Aachen, Germany}}
\newcommand{\SHEFFIELD}{\affiliation{University of Sheffield, Department of Physics and Astronomy, Sheffield, United Kingdom}}
\newcommand{\SKKU}{\affiliation{Sungkyunkwan University, Department of Physics, Suwon, Korea}}
\newcommand{\TOHOKU}{\affiliation{Tohoku University, Research Center for Neutrino Science, Sendai, Japan}}
\newcommand{\ERI}{\affiliation{University of Tokyo, Earthquake Research Institute, Tokyo, Japan}}
\newcommand{\KAMIOKA}{\affiliation{University of Tokyo, Institute for Cosmic Ray Research, Kamioka Observatory, Kamioka, Japan}}
\newcommand{\RCCN}{\affiliation{University of Tokyo, Institute for Cosmic Ray Research, Research Center for Cosmic Neutrinos, Kashiwa, Japan}}
\newcommand{\IPMU}{\affiliation{University of Tokyo, Kavli Institute for the Physics and Mathematics of the Universe (WPI), University of Tokyo Institutes for Advanced Study, Kashiwa, Japan}}
\newcommand{\TOKYO}{\affiliation{University of Tokyo, Department of Physics, Tokyo, Japan}}
\newcommand{\TODAI}{\affiliation{University of Tokyo, Tokyo, Japan}}
\newcommand{\TITECH}{\affiliation{Tokyo Institute of Technology, Department of Physics, Tokyo, Japan}}
\newcommand{\TRIUMF }{\affiliation{TRIUMF, Vancouver, British Columbia, Canada}}
\newcommand{\TORONTO}{\affiliation{University of Toronto, Department of Physics, Toronto, Ontario, Canada}}
\newcommand{\WARSAW}{\affiliation{University of Warsaw, Faculty of Physics, Warsaw, Poland}}
\newcommand{\WUT}{\affiliation{Warsaw University of Technology, Institute of Radioelectronics and Multimedia Technology, Warsaw, Poland}}
\newcommand{\WARWICK}{\affiliation{University of Warwick, Department of Physics, Coventry, United Kingdom}}
\newcommand{\WINNIPEG}{\affiliation{University of Winnipeg, Department of Physics, Winnipeg, Manitoba, Canada}}
\newcommand{\VT}{\affiliation{Virginia Tech, Center for Neutrino Physics, Blacksburg, Virginia, USA}}
\newcommand{\WROCLAW}{\affiliation{Wroclaw University, Faculty of Physics and Astronomy, Wroclaw, Poland}}
\newcommand{\YEREVAN}{\affiliation{Institute for Theoretical Physics and Modeling, Yerevan, Armenia}}
\newcommand{\YORK}{\affiliation{York University, Department of Physics and Astronomy, Toronto, Ontario, Canada}}
\newcommand{\KYIV}{\affiliation{Kyiv National University, Department of Nuclear  Physics, Kyiv, Ukraine}}
\newcommand{\YOKOHAMA}{\affiliation{Yokohama National University, Faculty of Engineering, Yokohama, Japan}}
\newcommand{\TUS}{\affiliation{Tokyo University of Science, Department of Physics, Chiba, Japan}}
\newcommand{\STOCKHOLM}{\affiliation{Stockholm University, Oskar Klein Centre and Department of Physics, Stockholm, Sweden}}
\newcommand{\STOCKHOLMA}{\affiliation{Stockholm University, Oskar Klein Centre and Department of Astronomy, Stockholm, Sweden}}
\newcommand{\IITG}{\affiliation{Indian Institute of Technology Guwahati, Guwahati, India}}
\newcommand{\IITK}{\affiliation{Indian Institute of Technology Kharagpur, Department of Physics, Kharagpur, India}}
\newcommand{\TEZPUR}{\affiliation{Tezpur University, Department of Physics, Sonitpur, India}}
\newcommand{\GLASGOW}{\affiliation{University of Glasgow, School of Physics and Astronomy, Glasgow, United Kingdom}}
\newcommand{\VICTORIA}{\affiliation{University of Victoria, Department of Physics and Astronomy, Victoria, British Columbia, Canada}}
\newcommand{\CARLETON}{\affiliation{Carleton University, Department of Physics, Ottawa, Ontario, Canada}}
\newcommand{\BCIT}{\affiliation{British Columbia Institute of Technology, Physics Department, Burnaby, British Columbia,  Canada}}
\newcommand{\AGH}{\affiliation{AGH University of Science and Technology, Faculty of Computer Science, Electronics and Telecommunications, Krakow, Poland}}
\newcommand{\OAUJ}{\affiliation{Astronomical Observatory of the Jagiellonian University, Krakow, Poland}}
\newcommand{\KEIO}{\affiliation{Keio University, Department of Physics, Yokohama, Japan}}
\newcommand{\CANFRANC}{\affiliation{Laboratorio Subterr\'aneo de Canfranc, Canfranc-Estaci\'on, Spain}}
\newcommand{\CRACOW}{\affiliation{H. Niewodnicza\'nski Institute of Nuclear Physics PAN, Cracow, Poland}}
\newcommand{\SILESIA}{\affiliation{University of Silesia in Katowice, A. Che\l{}kowski Institute of Physics, Poland}}
\newcommand{\KCL}{\affiliation{King's College London, Department of Physics, Strand Building, Strand, London, United Kingdom}}
\newcommand{\UU}{\affiliation{Uppsala University, Department of Physics and Astronomy, Uppsala, Sweden}}
\newcommand{\GUADALAJARA}{\affiliation{Universidad de Guadalajara, CUCEI,  Departamento de Fisica, Guadalajara, Jal., Mexico}}
\newcommand{\GUADALAJARAI}{\affiliation{Universidad de Guadalajara, CUCEA, IT.Ph.D. program, Guadalajara, Jal., Mexico}}
\newcommand{\SINALOA}{\affiliation{Universidad Autonoma de Sinaloa, Culiacan, Mexico}}
\newcommand{\MOSCOW}{\affiliation{Moscow State University, Department of Theoretical Physics, Moscow, Russia}}
\newcommand{\SALERNOA}{\affiliation{Universit\`a degli Studi di Salerno and INFN Gruppo Collegato di Salerno, Fisciano, Italy}}
\newcommand{\SALERNOB}{\affiliation{INFN Gruppo Collegato di Salerno, Fisciano, Italy}}
\newcommand{\KTH}{\affiliation{KTH Royal Institute of Technology, Department of Physics, Stockholm, Sweden}}
\newcommand{\LNL}{\affiliation{INFN Laboratori Nazionali di Legnaro, Legnaro (PD), Italy}}
\newcommand{\VIIT}{\affiliation{Vishwakarma Institute of Information Technology, Pune, India}}  
\newcommand{\CAMPANIA}{\affiliation{Universit\`a della Campania ''L. Vanvitelli'' and INFN Sezione di Napoli, Napoli, Italy}} 
\newcommand{\INFNNA}{\affiliation{INFN Sezione di Napoli, Napoli, Italy}}
\newcommand{\KAIST}{\affiliation{Korea Institute of Science and Technology, Department of Physics, Daejeon, Korea}}
\newcommand{\TMU}{\affiliation{Tokyo Metropolitan University, Department of Physics, Tokyo, Japan}}
\newcommand{\UNIST}{\affiliation{Ulsan National Institute of Science and Technology, Department of Physics, Ulsan, Korea}}
\newcommand{\KNU}{\affiliation{Kyungpook National University, Department of Physics, Daegu, Korea}}
\newcommand{\CHARLES}{\affiliation{Charles University, IPNP, FMF, Prague, Czech}}
\newcommand{\IITJ}{\affiliation{Indian Institute of Technology Jodhpur, Department of Physics, Karwar, Rajasthan, India}}
\newcommand{\ETHZ}{\affiliation{ETH Zurich, Institute for Particle and Astroparticle Physics, Zurich, Switzerland}}
\newcommand{\SANGYO}{\affiliation{Kyoto Sangyo University, Department of Astrophysics and Atmospheric Sciences, Kyoto, Japan}}
\newcommand{\ITESM}{\affiliation{Tecnologico de Monterrey, Escuela de Ingenieria y Ciencias, Zapopan, Jalisco, Mexico}}
\newcommand{\LPI}{\affiliation{P.N.Lebedev Physical Institute of the Russian Academy of Sciences, Moscow, Russia}}
\newcommand{\NNSO}{\affiliation{University of Tokyo, Next-generation Neutrino Science Organization, Kamioka, Japan}}
\newcommand{\MOMA}{\affiliation{University of Oviedo, Applied Mathematical Modeling Group/Department of Physics, Oviedo, Spain}}
\newcommand{\DIPC}{\affiliation{Donostia International Physics Center and Ikerbasque Foundation, Basque Country, Spain}}

\AGH
\BCIT
\CAMPANIA
\CARLETON
\UCI
\CHARLES
\CHONNAM
\DONGSHIN
\DIPC
\LLR
\EDINBURGH
\ETHZ
\GENEVA
\GIST
\GLASGOW
\GUADALAJARAI
\GUADALAJARA
\CRACOW
\IMPERIAL
\IITG
\IITJ
\IITK
\SALERNOB
\LNL
\BARI
\NAPOLI
\INFNNA
\PADOVA
\ROME
\INR
\YEREVAN
\SACLAY
\OAUJ
\JPARC
\KEIO
\KEK
\KCL
\KOBE
\KAIST
\KTH
\KYIV
\KYOTO
\SANGYO
\KNU
\LPNHE
\CANFRANC
\LANCASTER
\LIVERPOOL
\LSU
\MADRID
\MIYAGI
\ITESM
\MOSCOW
\NAGOYA
\STELAB
\KMI
\NCBJ
\OKAYAMA
\OCU
\MOMA
\OXFORD
\LPI
\RIO
\REGINA
\RWTH
\SALERNOA
\SHEFFIELD
\SILESIA
\SINALOA
\SOKENDAI
\RAL
\STOCKHOLMA
\STOCKHOLM
\SKKU
\TEZPUR
\TOHOKU
\TODAI
\ERI
\KAMIOKA
\RCCN
\IPMU
\NNSO
\TOKYO
\TITECH
\TMU
\TUS
\TORONTO
\TRIUMF
\UNIST
\UU
\VICTORIA
\VIIT
\VT
\WARSAW
\WUT
\WARWICK
\WINNIPEG
\WROCLAW
\YOKOHAMA
\YORK

\author{K.~Abe}\KAMIOKA\IPMU\NNSO
\author{P.~Adrich}\NCBJ
\author{H.~Aihara}\TOKYO\IPMU\NNSO
\author{R.~Akutsu}\TRIUMF
\author{I.~Alekseev}\LPI
\author{A.~Ali}\KYOTO
\author{F.~Ameli}\ROME
\author{L.H.V.~Anthony}\IMPERIAL
\author{A.~Araya}\ERI\NNSO
\author{Y. Asaoka}\KAMIOKA\NNSO
% ...

\collaboration{Hyper-Kamiokande Collaboration}

\date{\today}

\begin{abstract}

  Hyper-Kamiokande is the next generation underground water Cherenkov detector that builds on the highly successful Super-Kamiokande experiment. The detector which has an 8.4~times larger effective volume than its predecessor will be located along the T2K neutrino beamline and utilize an
  upgraded J-PARC beam with 2.6~times beam power. Hyper-K's low energy threshold combined with the very large fiducial volume make the detector unique, that is expected to acquire an unprecedented exposure of 3.8~Mton$\cdot$year over a period of 20~years of operation. Hyper-Kamiokande combines an extremely diverse science program including nucleon decays, long-baseline neutrino oscillations, atmospheric neutrinos, and neutrinos from astrophysical origins. The scientific scope of this program is highly complementary to liquid-argon detectors for example in sensitivity to nucleon decay channels or supernova detection modes.

  Hyper-Kamiokande construction has started in early 2020 and the experiment is expected to start operations in 2027. The Hyper-Kamiokande collaboration is presently being formed amongst groups from 19 countries including the United States, whose community has a long history of making significant contributions to the neutrino physics program in Japan. US physicists have played leading roles in the Kamiokande, Super-Kamiokande, EGADS, K2K, and T2K programs.
\end{abstract}

\maketitle


\newpage

\noindent
{\bf \large The Hyper-Kamiokande Experiment}
{\vskip 0.075in}

Hyper-Kamiokande (Hyper-K)~\cite{Abe:2018uyc} is based on the highly successful Super-Kamiokande (Super-K) detector and takes full advantage of a well-proven technology. The detector will be located at the Tochibora site, 8~km south of the Super-K detector site. The cavern has a rock overburden of 650~m, corresponding to 1,750 meters of water equivalent (m.w.e.). The detector tank is 71~m in height and 68~m in diameter with a total volume of $258$~kt. It is separated into an inner detector region containing 217~kt of water and an outer detector veto region. The inner detector is viewed by an array of 40,000 high QE Box \& Line (B\&L) PMTs with 50~cm diameter. The new PMT type has twice the photon detection efficiency compared to Super-K PMTs and with a better timing resolution of 2.6~ns (FWHM), and charge resolution of 30\%. The dark rate is reduced comparable to the 6~kHz at Super-K. 

Hyper-K will receive an intense neutrino beam from J-PARC at an off-axis angle of $2.5^{\circ}$. The beam power of the 30~GeV proton beam from the J-PARC Main Ring synchrotron will be upgraded from $515$~kW (currently achieved for T2K) to 1.3~MW by increasing the number of protons per beam pulse and a higher repetition rate. The J-PARC beam power upgrade proceeds in stages and will be completed in 2028. Similar to T2K a $\nu_\mu$ and $\bar{\nu}_\mu$ flux will be produced.  

At the beginning of Hyper-K, three near detectors will precisely characterize the neutrino beam at J-PARC. The existing {\it ND280} detector will be upgraded~\cite{Abe:2019whr} with fine-granularity plastic scintillator detectors, high angle TPC, and TOF detectors. With a larger angular acceptance and the 3D reconstruction capability of the new fine-granularity detector a significant reduction in the systematic uncertainty for oscillation analyses is expected.
An {\it Intermediate Water Cherenkov Detector} (IWCD)~\cite{Abe:2018uyc} is proposed to be newly constructed at about 1~km distance from the neutrino production target. This one kiloton-scale water Cherenkov detector can be moved vertically to measure the neutrino beam intensity and energy spectrum at different off-axis angles. The main science goals include neutrino cross section measurements (3\% for $\sigma(\nu_e)/\sigma(\nu_\mu)$ and 5\% for $\sigma(\bar{\nu}_e)/\sigma(\bar{\nu}_\mu)$) and measurement of intrinsic electron neutrino backgrounds.
The T2K on-axis near detector, {\it Interactive Neutrino GRID} (INGRID)~\cite{Abe:2015zbg}, monitors neutrino event rates and measures the neutrino beam direction with a precision better than 1~mrad.
INGRID is located at 280~m downstream from the production target will continue to be used for Hyper-K.

{\vskip 0.25in}
\noindent
{\bf \large Nucleon Decay, Dark Matter, BSM Physics}
{\vskip 0.075in}

\noindent
Hyper-K will be able to provide some of the most stringent tests of the Standard Model. 
Nucleon decay sensitivities will be extended by one order of magnitude beyond the current limits and could reveal grand unified theories (GUTs)~\cite{Abe:2018uyc}. With 20~years of data, Hyper-K will reach a proton decay sensitivity of $10^{35}$~years for $p \rightarrow \pi^0 e^+$ and $3\times 10^{34}$~years for $p \rightarrow \bar{\nu} K^+$. These decay modes are highly complementary to those accessible by liquid argon based detectors.

The search for physics beyond the Standard Model of particle physics is one of the priorities of Hyper-K. Following Super-K's success the indirect search for dark matter from the Sun is expected to continue to provide the most sensitive tests for dark matter nucleon scattering for masses of a few a GeV in a model independent way~\cite{Choi:2013eda,Choi:2015ara,Danninger:2014xza}. Hyper-K can search for boosted dark matter~\cite{Kachulis:2017nci} or test models with predominantly hadronic annihilation channels that remain hidden to other neutrino detectors~\cite{Rott:2012qb,Bernal:2012qh,Rott:2015nma}.
Sensitivity of searches for dark matter from the Galactic halo will provide some of the best sensitivities to dark matter for annihilation channels with large neutrino yields. 
Precise oscillation measurements with atmospheric neutrinos can be used to push limits for Lorentz symmetry violation~\cite{Abe:2014wla,Aartsen:2017ibm}, non-standard interaction~\cite{Mitsuka:2011ty,Aartsen:2017xtt}, quantum decoherence~\cite{Stuttard:2020qfv}, sterile neutrino oscillation~\cite{Abe:2019fyx,Aartsen:2020fwb}, and various dark sector particle searches such as  heavy neutral lepton~\cite{Abe:2019kgx,Coloma:2019htx}, long lived particle~\cite{Arguelles:2019ziu}, and
millicharged particles~\cite{Plestid:2020kdm}. Non-standard neutrino interactions can be probed by Hyper-K using the neutrino beam and could be enhanced with a second detector in Korea~\cite{Abe:2016ero,Liao:2016orc}.

{\vskip 0.25in}
\noindent
{\bf \large Solar Neutrinos}
{\vskip 0.075in}

Hyper-K, with its unprecedented statistical power, will be able to measure short-period 
flux variations in solar neutrinos, realizing a real-time monitoring of the Solar core temperature. Hyper-K could also achieve the first measurement of hep solar neutrinos, providing new insights in solar physics. The upturn at the vacuum-MSW transition region will be detectable at $5\sigma$ ($3\sigma$) with 10~years of data with an energy threshold of $3.5$~MeV ($4.5$~MeV).

{\vskip 0.25in}
\noindent
{\bf \large Astrophysics: Supernova Neutrinos, Multi-messenger science}
{\vskip 0.075in}

Through the observation of $\sim$10~MeV neutrinos with time, energy and directional information, Hyper-K will take a unique role as multi-messenger observatory. Hyper-K will expand on the successful multi-messenger science program of Super-K~\cite{Abe:2016jwn}. It has the potential to detect thermal neutrinos from nearby ($<10$~Mpc) neutron star merger events in coincidence with gravitational waves.
 
 Hyper-K provides unique sensitivity to core-collapse supernova neutrinos~\cite{Hirata:1987hu,Ikeda:2007sa,Nakamura:2016kkl} and is expected to detect more than 50,000~events for a Galactic supernova at 10~kpc.  Hyper-K's reach extends to the Andromeda Galaxy~M31 ($\sim 780$~kpc) with about 10 to 16 events expected per supernova. 
The direction of a supernova  at  10~kpc  can  be  reconstructed  with  an  accuracy  of  about  $1^\circ$  to  $1.3^\circ$ assuming similar event reconstruction performance as Super-K~\cite{Abe:2018uyc}, making Hyper-K essential for distributing early alerts and multi-messenger observations. Further, even a few neutrinos from nearby extra-galactic supernovae can reveal the nature of transients whose mechanism is uncertain~\cite{Thompson:2008sv}.

The search for diffuse supernova neutrino background (DSNB)~\cite{Beacom:2010kk,Lunardini:2010ab} will aid our understanding of the rate and spectrum of typical supernova explosions. With the addition of Gd~\cite{Beacom:2003nk}, Hyper-K could also separate $\bar{\nu}_e$ inverse beta reactions from other interactions. With Gd doping to tag $\bar{\nu}_e$ between 10 and 30~MeV and thereby distinguish supernova neutrinos from atmospheric neutrino backgrounds the sensitivity to DSNB can be significantly enhanced.

{\vskip 0.25in}
\noindent
{\bf \large CP Violation, Neutrino Mass Ordering, Non-standard interactions}
{\vskip 0.075in}

The apparent baryon asymmetry in our Universe is one of the greatest unsolved problems of our time. CP violation in the lepton sector could generate the observed matter-antimatter disparity through leptogenesis~\cite{Fukugita:1986hr}.  
In 2020, the T2K~\cite{Abe:2019vii} reported a measurement that favors large enhancement of the neutrino oscillation probability, excluding values of $\delta_{CP}$, which result in a large enhancement of the observed anti-neutrino oscillation  probability at $3\sigma$. The result follows previous preference in global fits~\cite{Esteban:2018azc} but is in tension with NO$\nu$A~\cite{deSalas:2020pgw}. 

With 10~years of running at 1.3~MW and split 1:3 between neutrinos and antineutrinos respectively, more than 65\% of $\delta_{CP}$ can be excluded at $3\sigma$ given $\sin \delta_{CP}=0$.  
A reduction in systematic uncertainties can further improve the sensitivity to $\delta_{CP}$. 

Using atmospheric neutrinos Hyper-K has sensitivity to determine the neutrino mass ordering due to the matter effects in the Earth. Sensitivity can be significantly enhanced with a detector in Korea~\cite{Abe:2018uyc}. The unitarity of the PMNS matrix can be tested through $\nu_{\tau}$ appearance~\cite{Parke:2015goa,Li:2017dbe,Aartsen:2019tjl}. These oscillation results relies on precise neutrino interaction measurements by new near detectors~\cite{Bhadra:2014oma,Sgalaberna:2017khy,Jones:2020dyx}. These new measurement will push current knowledge of neutrino-nucleus scattering physics~\cite{Alvarez-Ruso:2017oui}.

In a global context Hyper-K data combined with those from other long-baseline neutrino oscillation experiments will be essential to test non-standard oscillation scenarios, which can only be revealed by combining data from different experiments.

\vspace{-0.3cm}

\bibliographystyle{apsrev}
\bibliography{ref_rott}

\end{document}